\documentclass{appolb}
\usepackage{graphicx}
\usepackage{epsfig}
\def\Journal#1#2#3#4{{#1} {\bf #2}, #3 (#4)}

\def\NCI{\em Nuovo Cimento}

\def\NPB{{\em Nucl. Phys.} {\bf B}}

\def\NPA{{\em Nucl. Phys.} A}
\def\PLB{{\em Phys. Lett.}  {\bf B}}
\def\PRL{\em Phys. Rev. Lett.}

\def\PRC{{\em Phys. Rev.} C}

\def\PR{{\em Phys. Rev. }}
\def\PRP{{\em Phys. Rep. }}
\def\PPNP{{\em Prog. Part. Nucl. Phys. }}

\def\ZPA{{\em Z. Phys.} A}
\def\EPJA{{\em Europ. Phys. J.} A}
\def\EPJC{{\em Europ. Phys. J.} C}

\def\MPA{{\em Mod. Phys. Lett. A}}

\def\ARNP{{\em Ann. Rev. Nucl. Part. Phys.}}

\def\PTPS{\em Prog. Theo. Phys. Suppl.}

\def\RMP{\em Rev. Mod. Phys.}

\def\EPL{\em Europhys. Lett.}

\def\JPG{\em Journal of Physics G}

\def\ra{\rightarrow}
\def\be{\begin{equation}}
\def\ee{\end{equation}}

\newcommand{\expe}{experiment }

\newcommand{\exps}{experiments }

\newcommand{\bb}{double beta decay }
\newcommand{\obb}{0\mbox{$\nu\beta\beta$-decay} }
\newcommand{\zbb}{2\mbox{$\nu\beta\beta$-decay} }

\newcommand{\majo}{Majorana }
\newcommand{\nme}{nuclear matrix element }
\newcommand{\bdec}{$\beta$-decay }

\newcommand{\delm}{\mbox{$\Delta m^2$} }

\newcommand{\bpbp}{\mbox{$\beta^+\beta^+$} }
\newcommand{\ecec}{\mbox{$EC/EC$} }
\newcommand{\bec}{\mbox{$\beta^+/EC$} }

\newcommand{\bnel}{\mbox{$\bar{\nu}_e$} }

\newcommand{\nel}{\mbox{$\nu_e$}}
\newcommand{\nmu}{\mbox{$\nu_\mu$}}
\newcommand{\ntau}{\mbox{$\nu_\tau$}}

\newcommand{\ton}{\mbox{$T_{1/2}^{0\nu}$} }

\newcommand{\rh}{right--handed }
\newcommand{\neu}{neutrino }
\newcommand{\neus}{neutrinos }

\newcommand{\bea}{\begin{equation} \begin{array}{c}}
\newcommand{\eea}{ \end{array} \end{equation}}

\newcommand{\seza}{\mbox{$^{82}Se$ }}

\newcommand{\cdhs}{\mbox{$^{106}Cd$ }}

\newcommand{\tehd}{\mbox{$^{130}Te$ }}
 
\newcommand{\gess}{\mbox{$^{76}$Ge }}

\newcommand{\moeh}{\mbox{$^{100}$Mo }}

\newcommand{\cdhsz}{\mbox{$^{116}$Cd }}
\newcommand{\xehsd}{\mbox{$^{136}$Xe }}

\newcommand{\ema}{\mbox{$\langle m_{\nu_e} \rangle$ }}

\begin{document}
\title{Neutrinoless double beta decay experiments%
\thanks{Presented at Cracow Epiphany conference on Neutrinos and Dark 
Matter, 5.-8.1. 2006}%
}
\author{Kai Zuber
\address{Department of Physics and Astronomy, University of Sussex, \\
Falmer, Brighton BN1 9QH, UK}
}
\maketitle
\begin{abstract}
The study of neutrinoless double beta decay is of outmost importance for neutrino physics. It is considered to be the gold plated channel
to probe the fundamental character of neutrinos and to determine the neutrino mass. From the experimental point about nine
different isotopes are explored for the search. After a general introduction follows a short discussion on nuclear matrix element
calculations and supportive measurements.
The current experimental status of double beta searches is presented followed by a short discussion of the ideas and proposals for
large scale experiments.
\end{abstract}
\PACS{23.40-s, 14.60.Pq}
  
\section{Introduction}
\begin{figure}
\centering
\centering
\includegraphics[width=6cm]{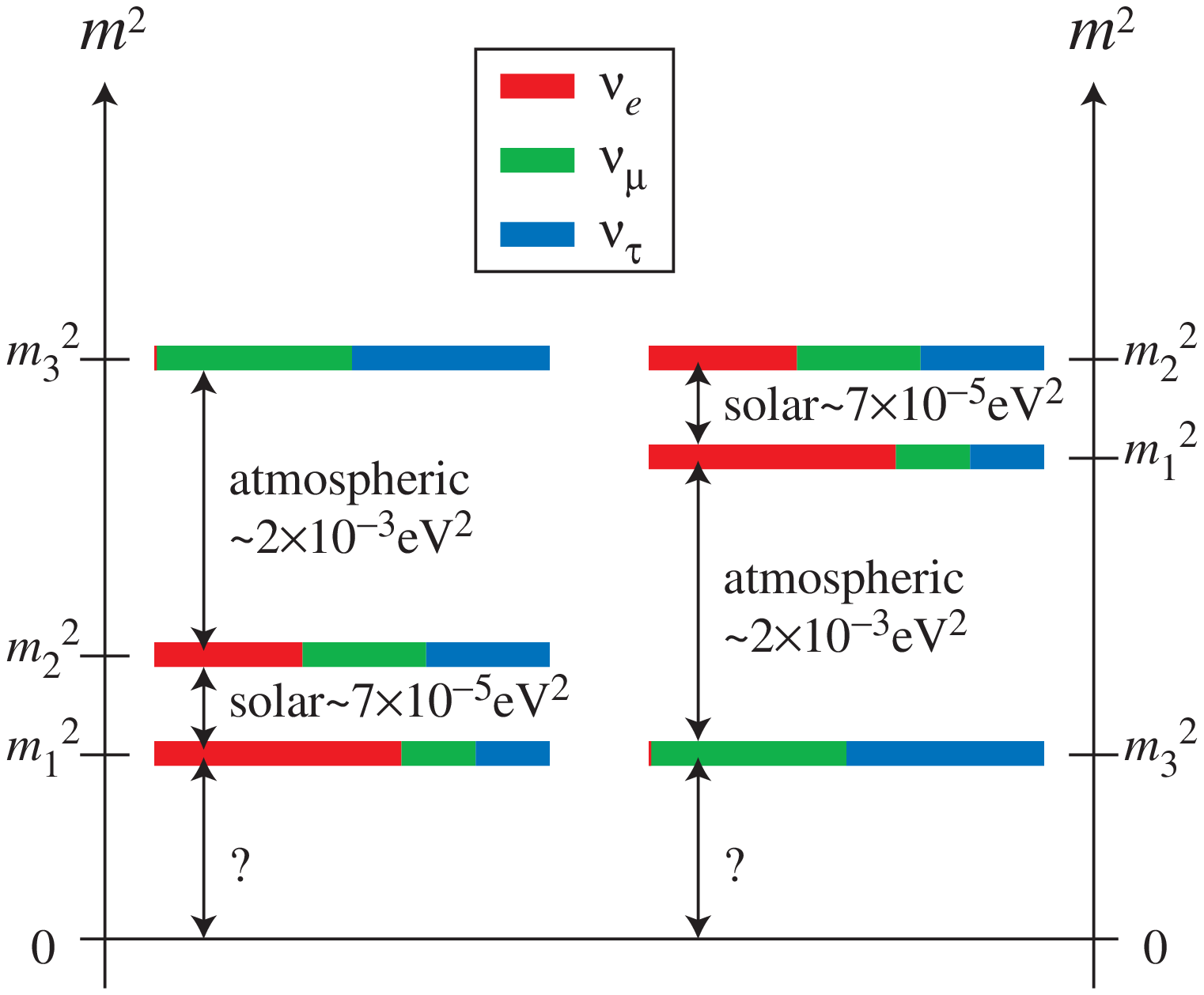}
\includegraphics[width=6cm]{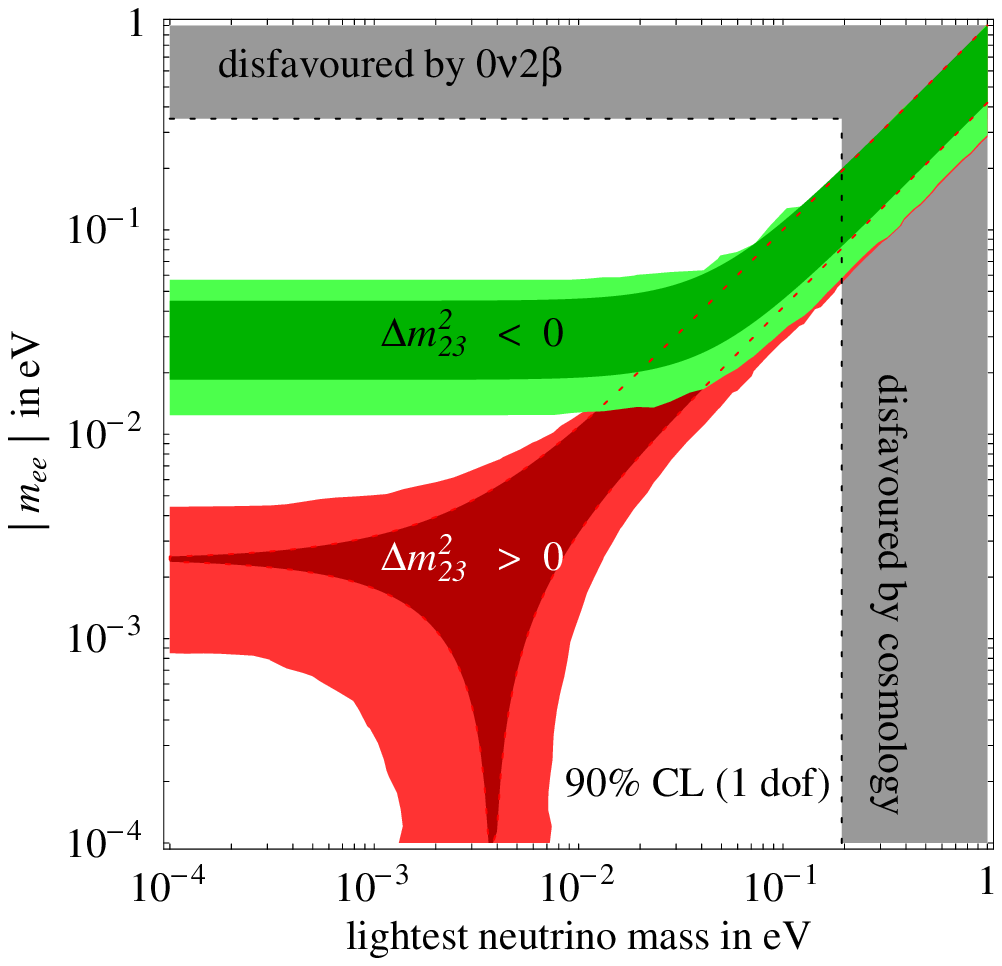}
\caption{\label{massm} Left: Possible configurations of neutrino mass states as suggested by oscillations. Currently
a normal (left) and an inverted (right) hierarchy cannot be distinguished.  The flavour composition is shown as well.
Right: The effective Majorana mass \ema as a function of the lightest mass eigenstate $m_1$. Hierarchical mass patterns
can be distinguished for \ema smaller than 50 meV, otherwise neutrinos can be considered as almost degenerate. Also
shown in grey are the regions disfavoured by current \obb limits and a very optimistic limit (could be worse by an order
of magnitude) from cosmology (from \protect \cite{fer02}).}
\end{figure}

Neutrino physics has gone through a revolution in the last ten
years. Now it is  beyond doubt that neutrinos have a non-vanishing rest mass. 
All the evidence stems from neutrino oscillation experiments, proving that neutrinos
can change their flavour if travelling from a source to a detector. 
Oscillations violate the concept of single lepton number conservation but total lepton number is still conserved.
Furthermore, the oscillation experiments
are not able to measure absolute neutrino masses, because their results depend only on the differences 
of masses-squared, \delm = $m_i^2- m_j^2$, with $m_i,m_j$ as the masses of two
neutrino mass eigenstates. In the full three neutrino mixing
framework the weak eigenstates \nel , \nmu\ and \ntau\ can be expressed as
superpositions of three neutrino mass eigenstates  $\nu_1, \nu_2$ and $\nu_3$ 
linked via a unitary matrix U:
\be
\label{mixingmatrix}
\left( \begin{array}{c}
\nel \\ \nmu \\ \ntau
\end{array}
\right) =
\left(
\begin{array}{ccc}
U_{e1} & U_{e2} & U_{e3}\\
U_{\mu 1} & U_{\mu 2} & U_{\mu 3}\\
U_{\tau 1} & U_{\tau 2} & U_{\tau 3}
\end{array}
\right)
\left( \begin{array}{c}
\nu_1 \\ \nu_2 \\ \nu_3
\end{array}
\right)
\ee
This kind of mixing has been known in the quark sector for decades and the analogous matrix 
U is called Cabbibo-Kobayashi-Maskawa matrix. The corresponding mixing matrix in the lepton sector
is named
Pontecorvo-Maki-Nakagawa-Sato (PMNS)-matrix. The unitary matrix U in eq. \ref{mixingmatrix} can 
be parametrised in the following form 
\be
\label{eq:ckmmatrix}
U = \left( \begin{array}{ccc}
c_{12} c_{13} & s_{12} c_{13} & s_{13} e^{- i \delta}\\
- s_{12} c_{23} - c_{12}s_{23}s_{13} e^{i \delta}& c_{12}c_{23} -
s_{12}s_{23}s_{13}e^{i \delta} &
s_{23}c_{13}\\
s_{12}s_{23}- c_{12}s_{23}s_{13} e^{i \delta} & -c_{12}s_{23} - s_{12}c_{23}s_{13}e^{i
\delta} &  c_{23}c_{13}
\end{array} \right)
\ee
where $s_{ij} = \sin \theta_{ij}, c_{ij} = \cos \theta_{ij}\: (i,j=1,2,3)$. The phase $\delta$ is a 
source for CP-violation and like in the quark sector cannot be removed by rephasing the neutrino
fields. 
The \majo case, ie. the requirement of particle and antiparticle to be identical, restricts the freedom to
redefine the fundamental fields even further. The net effect is the appearance of a CP-violating phase already in
two flavours. For three flavours two additional phases have to be introduced resulting in a mixing matrix
of the form
\begin{equation}
U = U_{PMNS} diag (1, e^{i \alpha_2}, e^{i \alpha_3})
\end{equation}
with the two new \majo phases $\alpha_2$ and $\alpha_3$. These phases again might only be accessible
in double beta decay, they are not accessible in neutrino oscillation experiments.
They are a further source of CP-violation.\\
Based on the observations from neutrino oscillations (see \cite{zub98,ose06}), various neutrino mass models have been proposed.
These can be categorized as normal hierarchy ($m_3 \gg m_2 \approx m_1$), 
inverted hierarchy ($m_2 \approx m_1 \gg m_3$)
and almost degenerate ($m_3 \approx  m_2 \approx m_1$) neutrinos (Fig. \ref{massm}). 
A key result, based on the observed $\Delta m^2$ in atmospheric neutrinos,
is the existence of a neutrino mass eigenstate in the region around 10-50 meV. This is the minimal
value neccessary, because it corresponds to the square root of the measured $\Delta m^2$ in case
one of the mass eigenstates is zero.
Fixing the absolute mass scale is of outmost importance, because it will fix the mixing matrix
and various other important quantities will then be determined, like the contribution of neutrinos to 
the mass density in the Universe. \\
Traditionally, laboratory experiments search for a finite neutrino rest mass by exploring
the endpoint energy of the electron spectrum in tritium beta decay. Currently a limit
for the electron neutrino mass of less than 2.2 eV has been achieved \cite{lob03,kra05}. A similar limit
is obtained by analysing recent cosmic microwave 
background measurements using the WMAP satellite combined with large scale
galaxy surveys and Lyman-$\alpha$ systems, see e.g. \cite{les06}. However, there are about two orders of magnitude
difference with respect to the region below 50 meV and even the next generation beta decay experiment, 
called KATRIN, can at best lead to an improvement
of a factor ten. However it should be noticed, that beta decay and \bb are measuring slightly different 
observables and are rather complementary than competitive.
Therefore, very likely double beta decay is the only way to explore the region below 100 meV.

\section{Double beta decay}
\begin{figure}
\centering
\includegraphics[height=4.5cm,width=9cm]{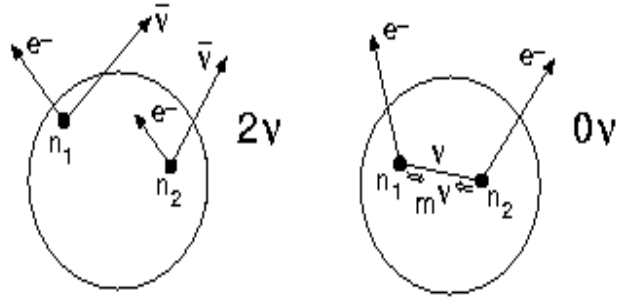}
\caption{\label{bbprinzip} Principle of double beta decay. Left: The simultaneous decay of two neutrons
as an allowed higher order process (\zbb). Right: The lepton-number violating mode (\obb) where the neutrino
only occurs as a virtual particle. This process is not allowed in the Standard Model.}
\end{figure}

Double beta decay is characterized by a nuclear process changing the nuclear charge Z by two 
units while leaving the atomic mass A unchanged. It is a transition among isobaric
isotopes.  It
is therefore a higher order process and can be seen as two simultaneous beta decays. This can only happen for 
even-even nuclei. All even-even nuclei have a ground state of spin 0 and
a positive parity, hence the ground state transitions are characterised as $(0^+ \ra 0^+)$ transitions.
Thus, a neccessary requirement for \bb to occur is $m(Z,A) > m(Z+2,A)$
and for practical purposes \bdec has to be forbidden $m(Z,A) < m(Z+1,A)$
or at least strongly suppressed. The same ground state configurations and arguments might hold for isotopes on the 
right side of the even-even parabola. This would lead to the process of double positron decay or double electron 
capture, discussed later.
In nature 35 isotopes are known, which show the specific ground state configuration, necessary for double beta decay.
Double beta decay was first discussed by M. Goeppert-Mayer \cite{goe35} in the form of 
\be
(Z,A) \ra (Z+2,A) + 2 e^- + 2 \bar{\nu_e} \quad (\zbb)
\ee
This process can be seen as two simultaneous neutron decays (Fig. \ref{bbprinzip}). 
Shortly after the classical papers of Majorana \cite{maj37} discussing a 2 - component
neutrino, Racah \cite{rac37} and Furry discussed another decay mode in form of \cite{fur39}
\be
\label{proc0nu}
(Z,A) \ra (Z+2,A) + 2 e^-  \quad (\obb) \quad .
\ee
In contrast to neutrino oscillations which violate individual flavour lepton number, but
keep total lepton number conserved, \obb\ violates total lepton number
by two units. This process is forbidden in the
Standard Model. It can be seen as two subsequent steps (''Racah - sequence'') 
as shown in Fig. \ref{bbprinzip}: 
\bea
(Z,A) \ra (Z+1,A) + e^- +  \bar{\nu_e} \nonumber\\
(Z+1,A) + \nel \ra (Z+2,A) + e^- \\ 
\eea
First a neutron decays under the emission of a
right-handed \bnel{}. This has to be absorbed at the second vertex as a left-handed \nel.
To fulfill these conditions neutrino and antineutrino have to be identical, requiring
that \neus are \majo particles, i.e. a 2-component object. This is different from all the other fundamental fermions
where particles and antiparticles can be already distinguished by their charge.
Majorana neutrinos are preferred by most Grand Unified Theories to explain the small magnitude of neutrino 
masses via the see-saw mechanism. Hence, double beta decay is generally considered to be the 'gold plated' channel 
to probe the fundamental character of neutrinos.
Moreover, to allow for the helicity matching
a \neu mass is required. The reason is that
the wavefunction describing neutrino mass eigenstates for $m_\nu >0$ has no fixed helicity
and therefore, besides the dominant left-handed contribution, has an
admixture of a right-handed component (or vice versa for antineutrinos), 
which is proportional to $m_\nu / E$.
Thus, for double beta decay to occur, massive \majo particles are required. For recent 
reviews on double beta decay see \cite{ell02,zde02a,ell04}.

The quantity measured in \obb is called effective
Majorana \neu mass and given for light neutrinos by 
\be 
\label{eq:ema}\ema = \mid \sum_i U_{ei}^2 m_i \mid =  \mid \sum_i \mid U_{ei} \mid^2 e^{2i\alpha_i} m_i \mid 
\ee 
which can be written in case of CP-invariance
as 
\be
\ema = \mid  m_1 \mid U_{e1}^2 \mid \pm m_2 \mid U_{e2}^2 \mid \pm m_3 \mid U_{e3}^2 \mid \mid
\ee 
As can be seen, the different terms in the sum have a chance to interfere destructively, only the absolute value
is measured at the end.
On the other hand, beta decay measures
\be
m_{\bnel} = \sum_i \mid U_{ei}^2 \mid m_i   
\ee
which is independent of the fundamental character of the neutrino and does not allow destructive interference.
As a result, a certain care should be taken if comparing \neu masses obtained
by \bdec and \obb, they should be seen as complementary measurements.

\section{General considerations}
\begin{figure}
\centering
\includegraphics[height=5cm,width=6.5cm]{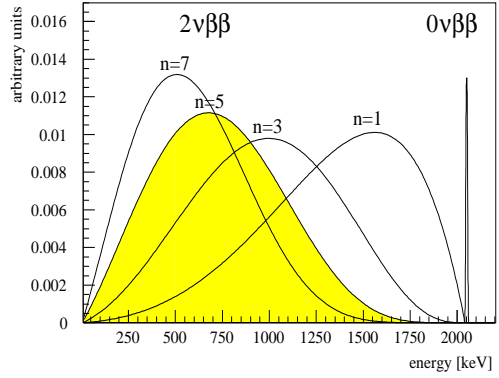}
\caption{\label{bbshape} Schematic drawing of the sum energy spectrum of electrons in double beta decay, here in 
case of $^{76}$Ge.The \zbb shows a continous spectrum (yellow), while \obb is a peak at the Q-value of the 
transition. The additional curves shown correspond to various majoron emitting modes not discussed here.}
\end{figure}

Being a nuclear decay, the actual experimental quantity measured is
the half-life. As a higher order effect the expected
half-lives for double beta decay are long, in the region of about $10^{20}$ years and beyond.
The experimental signal of \obb\ is two electrons in the final state, whose
energies add up to the Q-value of the nuclear transition, while for the \zbb
the sum energy spectrum of both electrons will be continuous (Fig.~\ref{bbshape}). The
total decay rates, and hence the inverse half-lives, are a strong function of the available Q-value. 
The rate of \obb scales with $Q^5$ compared to a $Q^{11}$-dependence for \zbb.
Therefore isotopes with a high Q-value (above about 2 MeV) are normally considered for experiments. 
This restricts one to eleven candidates listed in Tab.~\ref{tab:emitters}. 
The measured half-life or its lower limit in case of non-observation of the process
can be converted into a neutrino mass or an upper limit via 
\be
\label{eq:conversion}
(\ton )^{-1} = G^{0\nu} \mid M^{0\nu} \mid^2 \left( \frac{\ema}{m_e} \right)^2
\ee
where $ G^{0\nu}$ is the exactly calculable phase space integral (see \cite{boe92} for numerical
values) of the
decay and $\mid M^{0\nu} \mid$ is the nuclear matrix element
of the transition. 

With reasonable assumptions on the \nme it can be estimated that for a \neu
mass measurement of the order of 50 meV, half-lives in the region of $10^{26}-10^{27}$ years must
be explored, by no means an easy task. This can be shown by the following estimate:
Assume the radioactive decay law in the approximation $T_{1/2} \gg t$
\be
\ton = \ln 2 a  N_A m t/N_{\beta\beta}
\ee
with $t$ the measuring time, $m$ the used mass, $a$ is the natural abundance of the isotope of interest,
$N_A$ the Avogadro constant and
$N_{\beta\beta}$ the number of double beta decays. Expecting a half-life of about 6$\times 10^{26}$yrs and
to observe as little as one decay per year, the number of source atoms required is around 6 $\times 10^{26}$. This however 
corresponds to 1000 moles and using an average isotope of mass 100 like \moeh , would immediately imply using about 100 kg. 
Hence, even without any disturbing background, and full efficiency for detection, one needs about hundred
kilogram of the isotope of interest, to observe one decay per year independent of the experimental approach!
Even worse, in the background-limited case, 
the sensitivity on the half-life depends on experimental
quantities according to
\be
\label{hlfinneu}
\ton \propto a \cdot \epsilon \cdot \sqrt{\frac{M\cdot t}{\Delta E \cdot B}}
\ee
with $a$ the natural abundance of the isotope of interest, $\epsilon$ the 
detection efficiency, $M$ the mass of source employed, $t$ the
measuring time, $\Delta E$ the energy resolution at the peak position and $B$ the background index,
typically quoted in events/yr/keV/kg. In contrast to the background-free case, 
for a background-limited experiment 
the half-life sensitivity increases only with the square root of the measuring time and mass.

\begin{table}
\label{tab:emitters}
\caption{Compilation of $\beta^- \beta^-$-emitters with a Q-value of at least 2 MeV. Shown are
the transition energies Q and the natural abundances.}
\begin{center}
\begin{tabular}{@{}lll@{}}
\hline
Transition & Q-value (keV) & nat. ab. (\%) \\
\hline
$_{20}^{48}$Ca$\rightarrow _{22}^{48}$Ti&
4271  & 0.187 \\
$_{32}^{76}$Ge$\rightarrow _{34}^{76}$Se&
2039 & 7.8 \\
$_{34}^{82}$Se$\rightarrow _{36}^{82}$Kr&
2995  & 9.2 \\
$_{40}^{96}$Zr$\rightarrow _{42}^{96}$Mo&
3350 & 2.8 \\
$_{~42}^{100}$Mo$\rightarrow _{~44}^{100}$Ru&
3034 & 9.6 \\
$_{46}^{110}$Pd$\rightarrow _{48}^{110}$Cd&
2013 & 11.8 \\
$_{~48}^{116}$Cd$\rightarrow _{~50}^{116}$Sn&
2809 & 7.5 \\
$_{50}^{124}$Sn$\rightarrow _{52}^{124}$Te&
2288  & 5.64 \\
$_{~52}^{130}$Te$\rightarrow _{~54}^{130}$Xe&
2530  & 34.5\\
$_{~54}^{136}$Xe$\rightarrow _{~56}^{136}$Ba&
2479  & 8.9 \\
$_{~60}^{150}$Nd$\rightarrow _{~62}^{150}$Sm&
3367 & 5.6 \\
\hline
\end{tabular}
\end{center}
\end{table}

\subsection{Nuclear matrix elements}
\begin{figure} 
\centering
\includegraphics[width=10cm]{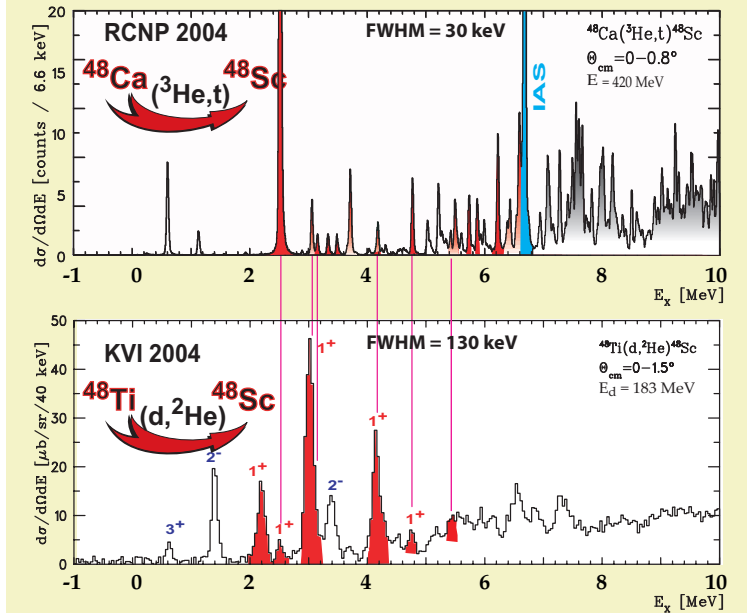}
\caption{\label{pic:GTTransitions} Measured $^{48}Ca(^3He,t)^{48}Sc$ (RCNP Osaka) and $^{48}Ti(d,^2He)^{48}Sc$ (KVI Groningen) 
spectra
in charge exchange reactions at 0 degree. Intermediate states which are excited by both reactions are 
on top of each other (from \protect \cite{fre06a}).}
\end{figure}

As can be seen in eq.\ref{eq:conversion} major ingredients in the conversion of
measured half-lives into neutrino masses are the involved nuclear matrix elements.
Those calculations are performed within the quasi random phase approximation (QRPA) or
by using the shell model. While \zbb matrix element are pure Gamow-Teller transitions as only $1^+$-states in the
intermediate nucleus are contributing, in \obb also higher multipoles contribute. A detailed discussion is beyond this article, for details see \cite{fae98,suh98,eji00,rod06}. There still seems to be an uncertainty of a factor 2-3 in the
calculations, the treatment of short range-correlation functions are likely responsible for a significant part of the 
discrepancy. 
Hence, an initiative has recently been started to provide those calculations with more and better
input from the experimental side to help as much as possible to settle the issue \cite{zub05}. 
Those measurement include charge exchange reactions measuring
the transition strengths to $1^+$-states. Some of the isotopes have already been measured at KVI Groningen with the (d,$^2$He)
reactions complemented by ($^3$He,t) measurements performed at RCNP Osaka (Fig.~\ref{pic:GTTransitions}). New ft-value measurements of electron capture for the intermediate nuclei are proposed using atomic traps \cite{fre06}. Those might help to solve the
issue of how to fix the partice-particle coupling parameter $g_{PP}$, to which the $1^+$-states calculations are very sensitive. Atomic traps will be of usage as well to determine the Q-values of some transitions more accurately
by high precision mass spectrometry. In addition, ordinary muon capture and neutrino-nucleus scattering have been proposed
to gain further information on the involved matrix elements.
The hope is that all those measurements might allow to bring down the error to the level of 30 \%. 

\section{Experimental status}
The search for \obb relies on finding a peak in the region below 4.3 MeV, depending on the isotope
(see Tab. \ref{tab:emitters}).
Common to all experimental approaches is 
the aim for a very low-background environment due to the fact of the expected long half-lives.
Among the most common background sources
are the natural decay chains of U and Th, $^{40}$K, Rn, neutrons, atmospheric muons and
radioisotopes produced in materials while on the surface. \\
All direct \exps are focusing on electron detection and can be either active or
passive.  
Active detectors are such that source and detector are identical which is a big advantage,
but often only measure the sum energy of both electrons. On the other hand, passive
detectors (source and detector are different) allow to get more
information like measuring energy and tracks of both electrons seperately, 
but usually have smaller source strength.
Some experiments will be described now in a little more detail.

\subsection{Ge-semiconductors - Heidelberg Moscow and IGEX}
\begin{figure}
\centering
\includegraphics[height=5cm,width=6.5cm]{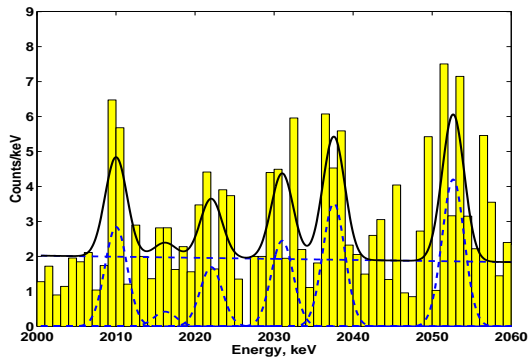}
\caption{\label{heimoevi} Energy spectrum of the Heidelberg-Moscow experiment around the 
\obb region at 2040 keV (from \protect \cite{kla04}).}
\end{figure}

The major progress in the last decades pushing half-life limits and increase the sensitivity towards
smaller and smaller neutrino masses have been achieved using Ge-semiconductor devices.
Source and detector are identical, the isotope under investigation is \gess
having a Q-value of 2039 keV.
The big advantage is the excellent energy resolution of Ge-semiconductors
(typically about 3-4 keV at 2 MeV). However, the technique only
allows the measurement of the sum energy of the two
electrons. A big step forward due to an increase in source strength was done by using enriched germanium 
(the natural abundance of \gess is 7.8 \%). Two experiments were performed recently, the Heidelberg-Moscow
and the IGEX experiment.
The Heidelberg-Moscow \expe in the Gran Sasso Laboratory took data from 1990-2003 using 
11 kg of Ge
enriched to about 86 \% in \gess in the form of five high purity Ge-detectors (HPGe).
A background as low as 0.12 counts/year/kg/keV at the peak position has 
been achieved. 
After 53.9 $kg\times y$ of data taking the peak region reveals no signal
and the obtained half-life limit is \cite{kla01}
$\ton > 1.9 \times 10^{25} $yrs$ (90 \% CL)$
which can be converted using eq.~\ref{hlfinneu} and the matrix elements given in 
\cite{sta90}
to an upper bound of $\ema < 0.35$ eV.
This is currently the best available bound coming from \bb. 
However, recently a subgroup of the collaboration found
a small peak at the expected position \cite{kla02,kla04} (Fig. \ref{heimoevi}).
Taking the peak as real and based on 71.7 $kg\times y$ of data would point towards a half-life between $0.7-4.2 \times 10^{25}$yrs.
Using the matrix elements calculated in \cite{sta90} this would imply a range for the neutrino
mass between 0.2-0.6 eV, which might be widened by using other matrix element calculations.
If true, this would immediately result in the fact that neutrinos are almost degenerate.
However, the discussion concerning the possible evidence is still quite controversial, 
see \cite{fer02,aal02,kla02a,har02,zde02}.

\subsection{CdZnTe-semiconductors - COBRA}
\begin{figure}
\centering
\includegraphics[height=4cm,width=6cm]{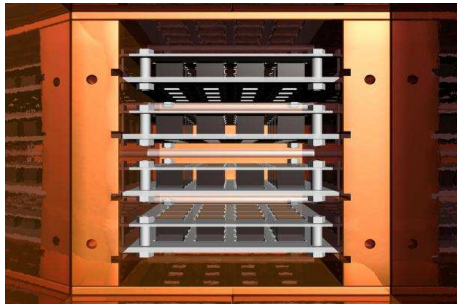}
\caption{\label{array} Schematic layout of the COBRA 64 array in form of a $4 \times 4 \times 4$ configuration.
Each layer contains 16 CdZnTe semiconductor detectors.}
\end{figure}

A new approach to take advantage of the good energy resolution
of semiconductors is COBRA \cite{zub01} located in the Gran Sasso Underground Laboratory
(LNGS). In total, there are seven (nine in case of CdZnTe) double
beta emitters within the detector including those of \bpbp decay.
The idea here is to use CdZnTe detectors, mainly to explore \cdhsz and
\tehd decay and \cdhs for \bpbp decay. The smallness of the
detectors makes a search for coincidences powerful and reduces $\gamma$-background. The practical
handling is simplified as these detectors are room temperature detectors. In case of 
pixelated detectors it offers
tracking possibilities and even further background reduction. 
Recent results obtained with four detectors can be found in \cite{zub06}.
Currently an upgrade to 64 detectors is ongoing, corresponding to about 0.42 kg CdZnTe (Fig.~\ref{array}). 

\subsection{Cryogenic bolometers - CUORICINO}
Currently only two large scale experiments are running.
The first technique uses bolometers
running at very low temperature (mK). CUORICINO at the Gran Sasso Underground Laboratory in Italy, 
is operating 62 TeO$_2$ crystals, corresponding to about 40 kg,  at 8 mK to search for the \tehd
decay with a Q-value of 2530 keV. The obtained half-life limit
corresponds to \cite{arn05} $\ton (\tehd) > 2.2 \times 10^{24} yr \quad (90 \% CL)$ resulting in a
upper bound on the neutrino mass of 0.2-1.1 eV, depending on the used matrix elements.

\subsection{Time projection chambers - NEMO-3}
The second experiment, NEMO-3 in the Frejus Underground Laboratory, is of the form of a passive \expe ,
which are mostly built in form of time
projection chambers (TPCs) where the double beta emitter is either the
filling gas of the chamber (like \xehsd ) or is included in thin foils. The advantage is that energy
measurements as well as tracking of the two electrons is possible. 
Disadvantages are the energy resolution and in case of using thin foils 
the limited source strength.
NEMO-3 consists of a tracking (wire chambers) and a calorimetric (plastic scintillators)
device put into a 25 G magnetic field. The total source strength is about 10 kg
which in a first run is dominated by using enriched $^{100}$Mo foils (about 7 kg).
Limits of $\ton (\moeh) > 5.6 \times 10^{23} yr \quad (90 \% CL)$ and $\ton (\seza) > 2.7 \times 10^{23} yr \quad (90 \% CL)$ 
resulting in upper neutrino mass bound of 0.6 - 2 eV from \moeh have been achieved \cite{sak06}.\\
Observation of \zbb has been quoted now for about a dozen isotopes. A complete listing of all experimental results obtained until end of 2001 can be
found in \cite{tre02}, some newer values are in \cite{ell04}, the most important ones are shown in Table~\ref{resultsbb}.

\begin{table}
\label{resultsbb}
\caption{Compilation of some obtained limits for \obb . However, notice the claimed evidence
for $^{76}$Ge. All results are 90 \% CL, except $^{48}$Ca (76 \%) and 
$^{128}$Te (68 \%).$^{\ast}$ corresponds to a geochemical measurement.}
\begin{center}
\begin{tabular}{@{}lll@{}}
\hline
Isotope & Half-life limit (yrs) & $\nu$ mass limit (eV)\\
\hline
$_{20}^{48}$Ca$\rightarrow _{22}^{48}$Ti&$>9.5\cdot10^{21} (76\%)$& $<8.3$\\
$_{32}^{76}$Ge$\rightarrow _{34}^{76}$Se&$>1.9\cdot10^{25} \hfill(90\%)$ & $<0.35 $ \\
$_{32}^{76}$Ge$\rightarrow _{34}^{76}$Se&$0.7-4.2 \cdot10^{25} \hfill(90\%)$ & $0.2-0.6 $ \\
$_{34}^{82}$Se$\rightarrow _{36}^{82}$Kr&$>2.7\cdot10^{23} \hfill(90\%)$ &$<5.0 $\\
$_{~42}^{100}$Mo$\rightarrow _{~44}^{100}$Ru&$>5.6\cdot10^{23} \hfill(90\%)$& $<0.6-2$ \\
$_{~48}^{116}$Cd$\rightarrow _{~50}^{116}$Sn&$>1.7\cdot10^{23} \hfill(90\%)$& $<1.7$ \\
$_{~52}^{128}$Te$\rightarrow _{~54}^{128}$Xe&$>7.7\cdot10^{24} \hfill(68\%)$& $< 1.1^{\ast}$\\   
$_{~52}^{130}$Te$\rightarrow _{~54}^{130}$Xe&$>2.2\cdot10^{24} \hfill(90\%)$& $<0.2-1.1 $\\   
$_{~54}^{136}$Xe$\rightarrow _{~56}^{136}$Ba&$>4.4 \cdot10^{23} \hfill(90\%)$& $<2.3$\\
$_{~60}^{150}$Nd$\rightarrow _{~62}^{150}$Sm&$>2.1\cdot10^{21} \hfill(90\%)$& $<4.1$ \\
\hline
\end{tabular}
\end{center}
\end{table}


\section{Interpretation of the obtained results}
 
It should be noted that double beta decay could also occur through other $\Delta L =2$ processes besides
light \majo \neu exchange. Whatever kind of new physics is providing this lepton number violation with two
electrons in the final state will be restricted by the obtained experimental results. Among them are right-handed
weak interactions, heavy \majo neutrino exchange, double charged higgs bosons, R-parity violating SUSY couplings, leptoquarks and other Beyond Standard
Model physics.\\
Assume that in addition to the neutrino mass mechanism,
also right handed leptonic and hadronic currents can contribute, i.e. the existence of a new (V+A)
interaction in addition to the well known (V-A) interaction.
The general Hamiltonian used for \obb rates is then given by
\be 
H = \frac{G_F \cos \theta_C}{\sqrt{2}} (j_L J_L^\dag + \kappa j_L J_R^\dag + 
\eta j_R J_L^\dag + \lambda j_R J_R^\dag)
\ee
with $G_F$ as the Fermi constant, $\cos \theta_C$ as the Cabibbo angle 
and the left- and right-handed leptonic currents given as
\be
j^\mu_L = \bar{e} \gamma^\mu (1 - \gamma_5 ) \nu_{eL} \quad j^\mu_R = \bar{e} \gamma^\mu (1 +
\gamma_5)
\nu_{eR}
\ee 
respectively.
The coupling constants $\kappa,\eta,\lambda$ vanish in the standard model
and $\kappa = \eta$ in left-right symmetric theories.
Expression \ref{hlfinneu} can be generalised if \rh currents are included to 
\begin{eqnarray}
\label{eq:cmm} (\ton)^{-1} & = & C_{mm} (\frac{\ema}{m_e})^2 + C_{\eta\eta} \langle \eta
\rangle^2 \\
& & + C_{\lambda \lambda} \langle \lambda \rangle ^2 
+ C_{m\eta}(\frac{\ema}{m_e})\langle \eta \rangle \nonumber \\
& & + C_{m\lambda}(\frac{\ema}{m_e})\langle\lambda\rangle 
+ C_{\eta\lambda}\langle \eta \rangle \langle \lambda \rangle  \nonumber
\end{eqnarray}
where the coefficients $C$ contain the phase space factors and the matrix elements
and the effective quantities are
\be 
\langle \eta \rangle = \eta \sum_j U_{ej}V_{ej} \quad \langle \lambda \rangle = \lambda \sum_j
U_{ej}V_{ej} 
\ee 
with $V_{ej}$ as the mixing matrix elements among the \rh \neu states.
Eq. \ref{eq:cmm} reduces to eq. \ref{hlfinneu} in case $\langle \eta \rangle,\langle \lambda
\rangle$ = 0.
Allowing also \rh currents to contribute, \ema is fixed by an ellipsoid. 
The weakest mass limit allowed occur for $\langle \lambda\rangle, \langle
\eta \rangle
\neq 0$. In this case the half-life limit of $^{76}$Ge corresponds to limits of
$\ema < 0.56~eV, 
\langle \eta \rangle < 6.5 \times 10^{-9}$ and
$\langle \lambda \rangle < 8.2 \times 10^{-7}$
respectively.

\subsection{$\beta^+\beta^+$-decay}
There is still more to investigate than the electron emitting 
double beta decay discussed. One example is the counterpart emitting two
positrons. Three different decay channels can be considered for the latter
\bea
(Z,A) \ra (Z-2,A) + 2 e^+ + (2 \nel) \\
e^- + (Z,A) \ra (Z-2,A) + e^+ + (2 \nel) \\ 
2 e^- + (Z,A) \ra (Z-2,A) + (2 \nel) 
\eea
where the last two cases involve electron capture (EC).
Especially the \bec mode shows an enhanced sensitivity to right handed weak currents 
\cite{hir94}.
The experimental signatures of the decay modes 
involving positrons in the final
state are promising because of two or four 511 keV photons.
Despite this nice signature, they are less often discussed in literature, because for each generated 
positron the available Q-value is reduced by 2 $m_ec^2$, which leads to much smaller decay rates than
in comparable \obb . Hence, for \bpbp{}-decay to occur, the Q-values
must be at least 2048 keV. Only six isotopes are know to have such a high Q-value, see Tab.~\ref{tab:bpbpemitters}. 
The full Q-value is only available in the \ecec mode.
Its detection is
experimentally more challenging, basically requiring the concept
of source equal to detector again. 
In the $0\nu$ mode  because of energy and momentum conservation additional
particles must be emitted like an $e^+ e^-$ pair or
internal bremsstrahlung photons. There will be a resonance enhancement in the decay rate if the initial and final states are
degenerate as has recently been explore in the context of radiative \ecec \cite{suj04}. 
Current half-life limits are 
of the order of 10$^{20}$ yrs obtained with $^{106}$Cd and $^{78}$Kr
for the modes involving positrons \cite{tre02}. 
The $^{106}$Cd system is currently explored by TGV2 \cite{ste05} and COBRA.
The COBRA \expe has the chance of simultaneously measuring 5
different isotopes for this decay channels \cite{zub01}. As the 
decay is intrinsic to the CdZnTe detectors one has a good chance to observe
the $2\nu$ \ecec{} and for the positron emitting modes coincidences among
the crystals can be used. 

\begin{table}
\label{tab:bpbpemitters}
\caption{Compilation of $\beta^+ \beta^+$-emitters requiring a Q-value of at least 2048 MeV. Shown are
the full transition energies Q-$4m_ec^2$ and the natural abundances.}
\begin{center}
\begin{tabular}{@{}lll@{}}
\hline
Transition & Q-$4m_ec^2$ (keV) & nat. ab. (\%) \\
\hline
$_{36}^{78}$Kr$\rightarrow _{34}^{78}$Se&
838 & 0.35 \\
$_{44}^{96}$Ru$\rightarrow _{42}^{96}$Mo&
676 & 5.5 \\
$_{~48}^{106}$Cd$\rightarrow _{~46}^{106}$Pd&
738 & 1.25 \\
$_{54}^{124}$Xe$\rightarrow _{52}^{124}$Te&
822  & 0.10 \\
$_{56}^{130}$Ba$\rightarrow _{54}^{130}$Xe&
534  & 0.11 \\
$_{~58}^{136}$Ce$\rightarrow _{~56}^{136}$Ba&
362  & 0.19 \\
\hline
\end{tabular}
\end{center}
\end{table}

\section{Future}
The future activities are basically driven by three factors: 
\begin{itemize}
\item Explore the claimed evidence observed for \gess
\item Increase sensitivity for neutrino masses down to 50 meV 
\item Explore further processes to disentangle the various underlying physics 
mechanisms discussed for neutrinoless double beta decay and to compensate for the nuclear matrix elements uncertainties.
\end{itemize}
To address the first topic, experiments have to come up with similar
good experimental parameters like the Heidelberg-Moscow experiment, ie.
about 10 yrs measuring time, 11 kg of high isotopical abundance (88 \%),
superb energy resolution and excellent low background in the peak region.
Partly those parameters can be compensated by using an isotope with higher Q-value and more favourable matrix elements. As shown in the previous section,
CUORICINO and NEMO-3 are starting to restrict the claimed region of neutrino
masses. Very likely the next experiment to join is GERDA \cite{woi06}, using the former
Heidelberg-Moscow and IGEX Ge-semiconductor detectors. Concerning the second
item various ideas and proposals are available which are listed in
Table.~\ref{tab:futureexp}. The last item requires the study of other processes like \bec modes, transitions to excited $2^+$-states \cite{hir94,doi85} or LFV processes using charged leptons
like $\mu \ra e + \gamma$ \cite{vog05}. Furthermore, to account for the possible physics processes
and matrix element uncertainties, the measurement of at least 3-4 different double beta isotopes
might be necessary \cite{sim05}. 

\begin{table}
\label{tab:futureexp}
\caption{Compilation of proposals for future experiments. This table is a slightly modified version of 
the one given in \cite{ell04} and does not claim to be complete.}
\begin{center}
\begin{tabular}{@{}lll@{}}
\hline
Experiment & Isotope & Experimental approach\\
\hline
CANDLES & ${48}$Ca & Several tons of CaF$_2$ crystals in Liquid scintillator \\
CARVEL & $^{48}$Ca & 100 kg $^{48}CaWO_4$ crystal scintillators \\
COBRA & $^{116}$Cd, \tehd & 420 kg CdZnTe semiconductors \\
CUORE & \tehd & 750 kg TeO$_2$ cryogenic bolometers \\
DCBA & $^{150}$Nd & 20 kg Nd layers between tracking chambers \\
EXO & $^{136}$Xe & 1 ton Xe TPC (gas or liquid) \\
GERDA & $^{76}$Ge & $\sim$ 40 kg Ge diodes in LN$_2$, expand to larger masses \\
GSO & $^{160}$Gd & 2t $Gd_2SiO_3:Ce$ crystal scint. in liquid scintillator\\
MAJORANA & $^{76}$Ge & $\sim$ 180 kg Ge diodes, expand to larger masses \\
MOON & \moeh & several tons of Mo sheets between scintillator \\
SNO++ & $^{150}$Nd & 1000 t of Nd-loaded liquid scintillator \\
SuperNEMO & $^{82}$Se & 100 kg of Se foils between TPCs \\
Xe & $^{136}$Xe & 1.56 t of Xe in liquid scintillator \\
XMASS & $^{136}$Xe & 10 t of liquid Xe \\
\hline
\end{tabular}
\end{center}
\end{table}

\section{Summary}
While the observed \zbb decay is the rarest processes ever observed, there is an enormous
physics potential in the lepton number violating process of \obb. In addition to the standard 
analysis, assuming the exchange of a light \majo \neu, various other kinds of $\Delta L =2$ can severely be restricted. There is a hotly debated evidence
for a signal in agreement with neutrino masses between 0.2-0.6 eV which would imply almost degenerate 
neutrinos. If this turns out not to be real, the next benchmark number experiments are aiming for, is the
50 meV range, implying hundreds of kilogram of material. After identifying a positive signal, it will be necessary to figure out which lepton number violating physics process is dominating neutrinoless double decay and especially the contribution of 
light Majorana neutrino exchange. Covering also the nuclear matrix uncertainties it will be necessary
to study several isotopes. Various experimental approaches are discussed to accomodate for this. 
New co-ordinated actions are on their way to provide the nuclear matrix element calculations
with better experimental input parameters.
Last, but not least, \obb might
be the only opportunity to access two further possible CP-violating phases associated with the \majo character
of the neutrino. This might be important in the context of leptogenesis, explaining the observed baryon asymmetry in the
Universe with the help of CP-violation in the lepton sector.

\end{document}